\documentclass[sigconf]{acmart}
\usepackage{multirow}
\usepackage{booktabs}   
\usepackage{hhline}    
\usepackage{makecell}

\usepackage{graphicx}
\usepackage{caption}

\usepackage{amsmath, amsfonts}
\usepackage{balance}

\usepackage{algorithm}
\usepackage{algorithmic}

\usepackage{balance}   
\usepackage{enumitem}   
\usepackage{xcolor}     
\usepackage{textcomp} 
\usepackage{url}        
\usepackage{verbatim}  
\usepackage{listings} 

\AtBeginDocument{%
  }

\copyrightyear{2026}
\acmYear{2026}
\setcopyright{cc}
\setcctype{by}
\acmConference[WWW '26]{Proceedings of the ACM Web Conference 2026}{April 13--17, 2026}{Dubai, United Arab Emirates}
\acmBooktitle{Proceedings of the ACM Web Conference 2026 (WWW '26), April 13--17, 2026, Dubai, United Arab Emirates}
\acmPrice{}
\acmDOI{10.1145/3774904.3792094}
\acmISBN{979-8-4007-2307-0/2026/04}

\begin{document}
% \settopmatter{printacmref=false}
% \settopmatter{printacmref=false, printccs=false, printfolios=false}
% \renewcommand\footnotetextcopyrightpermission[1]{}
% % \renewcommand\footnotetextcopyrightpermission[1]{}
% \pagestyle{plain}

\title{Beyond Patches: Superpixel Token-based Transformers for Attribute-Specific Fashion Retrieval}

\author{Shuili Zhang}
\authornote{Equal contribution.}
\author{Hongzhang Mu}
\authornotemark[1]
\affiliation{
  \institution{Institute of Information Engineering, Chinese Academy of Sciences \\School of Cyber Security, UCAS$^\ddagger$}
  \city{Beijing, China} \\
  \country{zhangshuili@iie.ac.cn}
  \country{muhongzhang@iie.ac.cn}
 }

\author{Wenyuan Zhang}
\affiliation{
    \institution{Institute of Information Engineering, Chinese Academy of Sciences \\School of Cyber Security, UCAS$^\ddagger$}
  \city{Beijing, China} \\
  \country{zhangwenyuan@iie.ac.cn}
 }

\author{Duohe Ma}
\authornote{Corresponding author.\\ $^\ddagger$University of Chinese Academy of Sciences.}
\author{Tingwen Liu}
\authornotemark[2]
\affiliation{
    \institution{Institute of Information Engineering, Chinese Academy of Sciences \\School of Cyber Security, UCAS$^\ddagger$}
  \city{Beijing, China} \\
  \country{maduohe@iie.ac.cn}
  \country{liutingwen@iie.ac.cn}
 }

\renewcommand{\shorttitle}{Beyond Patches: Superpixel Token-based Transformers for Attribute-Specific Fashion Retrieval}
\renewcommand{\shortauthors}{Shuili Zhang, Hongzhang Mu, Wenyuan Zhang, Duohe Ma, \& Tingwen Liu}

\begin{abstract}
 Attribute-Specific Fashion Retrieval (ASFR) aims to improve fine-grained image retrieval by focusing on specific attributes. However, existing patch-based attention and Transformer methods often misalign with irregular attribute regions and are prone to background noise, limiting their ability to capture subtle, pixel-level microstructures. To tackle these challenges, we propose \textbf{\textit{Super}Fashion}., the first ASFR framework that adopts superpixel tokens within a Transformer architecture. \textit{Super}Fashion initially employs an attribute-guided attention mechanism to extract attribute-related features, which in turn guide the cropping of semantically meaningful image regions. Superpixel segmentation is then leveraged on these regions to generate compact, semantically coherent superpixel tokens. By incorporating modality-specific embeddings for both attribute and superpixel tokens, the superpixel token-based Transformer facilitates adaptive interaction and fusion, thereby enhancing attribute localization and discrimination. Extensive experiments on FashionAI, DARN, and DeepFashion demonstrate relative overall MAP improvements of \textbf{1.84\%}, \textbf{9.27\%}, and \textbf{9.35\%} over prior SOTA. \textit{Super}Fashion offers a new solution for web-based image retrieval. 
\end{abstract}




\maketitle

\begin{figure}[t]
    \centering
\includegraphics{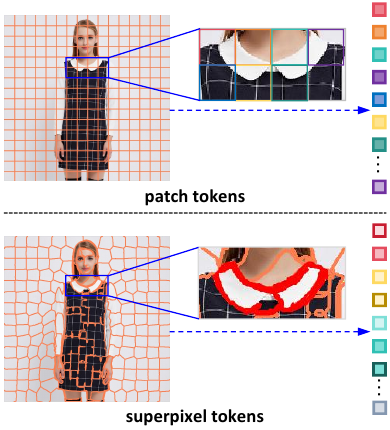}
\caption{Comparison of patch tokens and superpixel tokens for images with the same attribute: tokenization effects.}
    \label{fig:supertoken}
\end{figure}
\section{Introduction}
\label{sec:intro}
Fashion image retrieval \cite{tian2023FIR1,d2021FIR2} is a pivotal task in modern Web applications, particularly in the fashion domain, where users demand precise, highly attribute-aware search capabilities. However, conventional retrieval systems often rely on global visual similarity, which struggles to capture fine-grained attribute variations essential for fulfilling user intent (e.g., retrieving dresses with a specific neckline design). To address this, \textbf{A}ttribute-\textbf{S}pecific \textbf{F}ashion \textbf{R}etrieval (ASFR) has emerged as a paradigm that prioritizes precise attribute-level alignment over coarse global matching~\cite{fir_2021, song2022Con, zhang2023fashion, guo2023multimodal, Ma_ASEN_2020, Dong_RPF_2023}.
The demand for ASFR is especially high in web-based fashion applications and online shopping, as attribute-based retrieval \cite{DBLPZhangMLTS24} helps users quickly find items with specific features, such as a red handbag with chain straps, greatly improving product discoverability and streamlining the shopping experience. Likewise, in fashion communities and social platforms, ASFR empowers users to explore style variations or identify items with targeted attributes, fostering engagement and creative inspiration. Beyond enhancing retrieval accuracy, ASFR supports interpretable, user-controllable search experiences, aligning with the increasing focus on transparency and personalization in Web-scale systems~\cite{Veit_CSN_2017, Jiao_M3NET_2023, Jiao_MODC_2022}.

The core challenge of the ASFR task lies in accurately localizing attribute-aware features within images according to the specified attribute, and retrieving semantically visually diverse items that consistently manifest these characteristics. This is inherently difficult because attribute-specific cues vary substantially in form: attributes such as \texttt{neckline design} or \texttt{sleeve length} are confined to small, irregular regions, whereas others, such as \texttt{fabric} or \texttt{texture}, are distributed heterogeneously and may appear as fragmented, or subtle patterns across the image. To tackle these challenges, recent studies have explored attribute-guided attention mechanisms, designed to emphasize attribute-related regions and suppress irrelevant context \cite{Ma_ASEN_2020, yan2021learning, wan2024learning, ISLN_2022, Dong_ASEN++_2021}. Extending these studies, more recent approaches incorporate iterative attention refinement and attribute-aware transformers, thereby facilitating richer feature interactions and improving performance \cite{Dong_RPF_2023}.

Despite recent progress, existing studies still suffer from significant limitations that substantially constrain fine-grained retrieval. As shown in Figure~\ref{fig:supertoken}, patch-level attention mechanisms operate on uniformly partitioned image grids, which are inherently misaligned with the irregular shapes and diverse scales of attribute regions. This coarse partitioning prevents accurate modeling of subtle, pixel-level structures crucial for attribute discrimination. Moreover, patch-based regions frequently encompass irrelevant background content, introducing considerable noise and thereby diluting the distinctiveness of attribute-specific features. These deficiencies underscore a fundamental and persistent gap in current methods and highlight the urgent need for more adaptive and fine-grained solutions to achieve precise attribute localization.

Inspired by superpixel theory, we propose a novel framework, \textbf{\textit{Super}Fashion}, to address the inherent limitations of patch-based approaches. Notably, unlike conventional patch tokens, which suffer from rigid partitioning and susceptibility to background noise, \textit{Super}Fashion explicitly introduces \textbf{superpixel-level tokens} that naturally align with irregular attribute regions and preserve fine-grained structures. Specifically, the framework first employs an attribute-guided attention mechanism to extract attribute-related features. These features guide the cropping of image regions, ensuring that subsequent superpixel segmentation focuses on meaningful content. The cropped regions are processed through a screening structure to generate compact and semantically coherent superpixel tokens. The superpixel tokens and attribute tokens are individually augmented with modality-specific embeddings before being fed into the Transformer, enabling adaptive interaction and fusion while enhancing both attribute localization and discrimination.

In summary, the main contributions are summarized as follows:
\begin{itemize}
\item We propose a new approach for ASFR using superpixel tokens, effectively addressing misalignment and background noise while capturing fine-grained attribute structures.
\item We present \textit{Super}Fashion, the first framework to ingeniously generate superpixel tokens in a Transformer architecture for discriminative attribute-aware representations.
\item Extensive experiments on FashionAI, DARN, and DeepFashion demonstrate relative MAP improvements of \textbf{1.84\%}, \textbf{9.27\%}, and \textbf{9.35\%} over state-of-the-art baseline models.
\end{itemize}

\section{Related Work}
\label{sec:relatedwork}
\subsection{Attribute-Specific Fashion Retrieval}
In recent years, attribute-specific fashion retrieval has received growing attention in both academia and industry \cite{ISLN_2022, han2023fashionsap, Jiao_M3NET_2023, wan2024learning}. Early methods focused on extracting attribute-relevant regions via attention mechanisms. For example, CSNs \cite{Veit_CSN_2017} employed fixed masks to select attribute-specific embedding dimensions from global features, enabling fine-grained similarity measurement. ASEN \cite{Ma_ASEN_2020} further introduced Attribute-aware Spatial Attention (ASA) and Attribute-aware Channel Attention (ACA) to jointly learn multiple attribute embeddings in an end-to-end manner. Subsequent studies extended these mechanisms, including hierarchical attribute embeddings \cite{yan2021learning} and parallel ASA/ACA modules \cite{wan2024learning}, yet the coarse segmentation inherent in these region-based methods often introduces background noise and limits localization precision.
To improve granularity, patch-based strategies were proposed. Dong et al. \cite{Dong_ASEN++_2021} extracted patch-level features through repeated applications of ASA and ACA, while RPF \cite{Dong_RPF_2023} combined attention-guided patch extraction with Transformer architectures for enhanced attribute localization. Despite reducing noise, these approaches rely on fixed patch partitions, restricting adaptability to diverse attribute shapes and scales and resulting in imprecise boundaries and incomplete feature capture.
Recent advances have incorporated complementary techniques such as contrastive learning and knowledge distillation. Methods leveraging weak geometric distortion constraints \cite{xiao2025geodcl} or relational knowledge distillation \cite{xiao2024boosting} have achieved notable performance gains and enhanced industrial applicability.
In summary, while patch-level attention improves attribute-related retrieval, its coarse, grid-based partitioning remains misaligned with irregular attribute regions, limiting fine modeling and introducing background noise. These challenges highlight the need for adaptive and semantically coherent tokenization strategies, which we address.

\subsection{Visual Tokenization}

Most vision Transformer variants have focused on enhancing backbone architectures and attention mechanisms based on square image patches. Recently, research focus has shifted toward more advanced tokenization strategies that adapt dynamically to image content. For instance, Quadformer \cite{ronen2023quadFormer} and MSViT \cite{havtorn2023msvit} introduce adaptive tokenization schemes that dynamically adjust token resolution according to local image structures. SPiT \cite{aasan2024spit} applies superpixel-based tokenization; however, its primary focus is interpretability rather than performance improvement, and its conversion of superpixels into square patches can distort object structures. Other approaches, such as VCT \cite{yang2022vistoken2}, decompose images into unsupervised, disentangled visual concept tokens, while ViTok \cite{hansen2025vistoken1} employs autoencoding for latent tokenization in image and video generation, and TexTok \cite{zha2025vistoken3} constrains tokenization to descriptive captions to facilitate semantic learning. SuiT \cite{lew2025superpixeltokenizationvisiontransformers} introduces a superpixel-based tokenization method that replaces fixed grid patches in ViTs with adaptive superpixel tokens.
\begin{figure*}[t]
    \centering
\includegraphics[width=\textwidth]{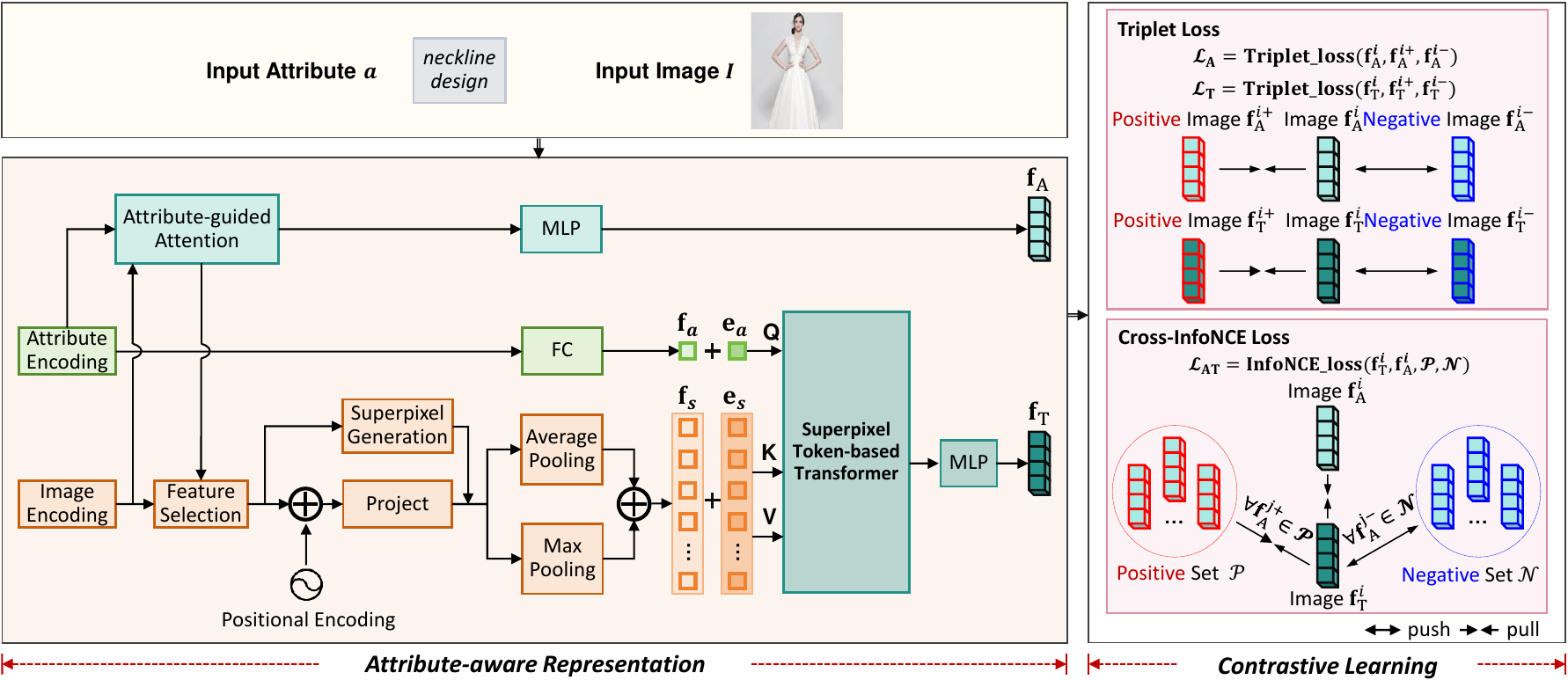}
    \caption{An overview of the proposed framework \textit{Super}Fashion, the two representations $\mathbf{f}_{\text{A}}, \mathbf{f}_{\text{T}}$ are used together for inference.}%, }
    \label{fig:framework}
\end{figure*}
Despite these recent advances, existing tokenization strategies for general vision tasks have not yet been systematically applied to ASFR tasks. In this paper, we specifically leverage carefully designed superpixel-based tokenization to enable precise attribute-aware partitioning, producing compact and semantically coherent tokens that effectively capture fine-grained, localized semantic regions, which are critical for achieving accurate and robust web-based attribute-specific fashion image retrieval.
\subsection{Superpixel Generation}
A superpixel is a cluster of homogeneous pixels defined by characteristics such as color, brightness, or texture \cite{chen2023deep}. As an over-segmentation technique, superpixels aggregate pixels into perceptually meaningful and semantically coherent regions, and they are widely used in computer vision applications \cite{liu2023superpixel}. Recent research has increasingly focused on improving boundary adherence, scalability, and adaptability for a variety of tasks, including semantic segmentation and object detection \cite{zhang2023superpixel, shen2016fastslic, guo2023superpixel}. Unsupervised and weakly supervised approaches aim to significantly reduce reliance on labeled data \cite{goyal2017unsupervised}, while deep learning, including CNNs and Transformer-based models, has been increasingly employed for adaptive, content-aware superpixel generation \cite{liu2021swin, chen2023multiscale}. Computational efficiency remains a key consideration, motivating lightweight and highly efficient designs suitable for real-time applications on resource-constrained devices \cite{zhang2023lightweight, xie2025novel, chen2023multiscale}. 
% Nonetheless, classical methods such as SLIC \cite{achanta2012slic} and FastSLIC \cite{shen2016fastslic} remain widely used owing to their simplicity, speed, and robust segmentation quality.
Nonetheless, classical methods such as SLIC \cite{achanta2012slic} and FastSLIC \cite{shen2016fastslic} remain widely used owing to their simplicity, speed, and consistent quality.

\section{Methodology}
\subsection{Overview of \textit{Super}Fashion Framework}
As illustrated in Figure~\ref{fig:framework}, \textit{Super}Fashion operates through three key stages. First, it generates superpixel tokens by using attribute-guided attention to extract features $\mathbf{f}_{\text{A}}$ for region cropping, then applies superpixel segmentation and aggregation to create semantically coherent tokens. Second, these tokens are processed through a superpixel token-based Transformer, producing refined representations $\mathbf{f}_{\text{T}}$ via an MLP. Finally, triplet loss and cross-InfoNCE loss are employed for contrastive learning, where the latter enables interactive learning between $\mathbf{f}_{\text{A}}$ and $\mathbf{f}_{\text{T}}$ to enhance joint inference.
\subsection{Superpixel Tokenization}
\subsubsection{Feature Extraction and Selection}
We first extract features from the input image $I$ using a convolutional block, resulting in a feature map $\mathbf{f}_{\text{I}}^{\text{o}} \in \mathbb{R}^{C \times H \times W}$. In parallel, the input attribute $a$ is encoded into an embedding vector $\mathbf{f}_a^{\text{o}} \in \mathbb{R}^A$ through an attribute encoding module. To facilitate cross-modal interaction, both image features and attribute embeddings are projected into a shared latent space: image features are transformed via a $1 \times 1$ convolutional layer, while attribute embeddings are processed with a fully connected (FC) layer. A subsequent $\tanh$ activation then yields the projected representations $\mathbf{f}_{\text{I}}^{\text{o}^\prime} \in \mathbf{R}^{C^\prime \times H^\prime \times W^\prime}$ and $\mathbf{f}_a^{\text{o}^\prime} \in \mathbf{R}^{C^\prime}$.  
To improve efficiency and focus on semantically relevant regions, we introduce an attribute-guided attention mechanism. This mechanism selectively emphasizes attribute-related patterns while suppressing background noise, motivated by the observation that attribute-specific cues are typically localized rather than globally distributed. The refined features $\mathbf{f}_{\text{A}}$ are obtained through the following operations:

\begin{equation}
    \boldsymbol{\alpha} = \text{softmax}(\mathbf{f}_{\text{I}}^{\text{o}'} \cdot \mathbf{f}_a^{\text{o}'}) \in \mathbb{R}^{H' \times W'},
\label{eq:attention_alpha}
\end{equation}
\begin{equation}
    \mathbf{f}_{\text{I}a} = \sum_{j}^{H' \times W'} \boldsymbol{\alpha}_j \mathbf{f}_{Ij}^{o},
\label{eq:weighted_sum}
\end{equation}
\begin{equation}
    \mathbf{f}_{\text{A}} = \text{LN}\left(\mathbf{W}_2 \left(\text{relu}\left(\mathbf{W}_1 \left(\text{LN}(\mathbf{f}_{\text{I}a})\right)\right)\right) + \mathbf{f}_{\text{I}a}\right),
\label{eq:residual_block}
\end{equation}
where $\mathbf{f}_{Ij}^{o}$ is the $j$-th channel-aware feature vector of $\mathbf{f}_{I}^{o}$, LN denotes layer normalization, and $\mathbf{W}_1$ and $\mathbf{W}_2$ are trainable weights.

\subsubsection{Superpixel-level Aggregation}
Following the attribute-guided attention mechanism, we crop the input image $I$ to obtain $I_c$, from which we extract image features $\mathbf{f}_\text{I} \in \mathbb{R}^{D_\text{I} \times H \times W}$ for superpixel token generation. Inspired by SuiT \cite{lew2025superpixeltokenizationvisiontransformers}, we propose superpixel feature aggregation into the tokenization process. To preserve high-frequency details, we adopt a superpixel-based positional encoding scheme. Specifically, sinusoidal positional encoding with learnable frequencies \cite{tancik2020fourier} is applied, yielding positional features $\mathbf{f}_\text{P} \in \mathbb{R}^{D_\text{P} \times H \times W}$ for each spatial location $(h, w)$ as:
\begin{equation}
    \mathbf{f}_{\text{P}}^{(h,w)}[2q] = \sin(g_x[q]\cdot h + g_y[q]\cdot w),
\end{equation}
\begin{equation}
    \mathbf{f}_{\text{P}}^{(h,w)}[2q+1] = \cos(g_x[q]\cdot h + g_y[q]\cdot w),
\end{equation}
where $g_x[q]$ and $g_y[q]$ represent learnable frequencies along the horizontal and vertical axes for the $q$-th dimension.
We then combine the attribute-related image features $\mathbf{f}_\text{I}$ and positional features $\mathbf{f}_{\text{P}}$ through channel-wise concatenation and linear projection:

\begin{equation}
    \mathbf{f}_{\text{IP}} = [\mathbf{f}_{\text{I}} \oplus \mathbf{f}_{\text{P}}]\mathbf{W}_{\text{IP}}, \quad \mathbf{f}_{\text{IP}} \in \mathbb{R}^{\frac{D}{2} \times H \times W},
\end{equation}
where $\mathbf{W}_{\text{IP}} \in \mathbb{R}^{(D_\text{I}+D_\text{P})\times \frac{D}{2}}$ denotes the projection matrix, and $\oplus$ represents concatenation.  
Given the superpixel index map $g_s$ obtained from the FastSLIC algorithm \cite{shen2016fastslic,achanta2012slic} applied to the cropped image $I_c$, pixel-level embeddings within each superpixel $C_k$ are aggregated using a dual-pooling strategy, through average and max pooling:
\begin{equation}
    \mathbf{f}_{\text{avg}}^{k} = \frac{1}{|C_k|} \sum_{x^{(h,w)} \in C_k} \mathbf{f}_{\text{IP}}^{(h,w)},
\end{equation}
\begin{equation}
    \mathbf{f}_{\text{max}}^{k} = \max_{x^{(h,w)} \in C_k} \mathbf{f}_{\text{IP}}^{(h,w)},
\end{equation}
where $|C_k|$ denotes the number of pixels in superpixel $C_k$. The final superpixel embedding combines the results of the dual-pooling:
\begin{equation}
    \mathbf{f}_s^{k} = \mathbf{f}_{\text{avg}}^{k} \oplus \mathbf{f}_{\text{max}}^{k}.
\end{equation}
This dual-pooling strategy enables the superpixel embeddings to capture both global context and prominent local details.
 %within each superpixel.
\subsection{Superpixel Token-based Transformer}

Building upon the extracted superpixel tokens $\mathbf{f}_s = [\mathbf{f}_s^1, \mathbf{f}_s^2, \dots, \mathbf{f}_s^n]$, where $\mathbf{f}_s^i \in \mathbb{R}^D$ denotes the embedding of the $i$-th token and $n$ is the total number of superpixel tokens, we employ a superpixel token-based Transformer module to capture attribute-aware features under attribute guidance. This module incorporates modality-specific embeddings $[\mathbf{e}_a, \mathbf{e}_s]$ to differentiate attribute tokens from visual tokens, encoding modality-specific information into their respective spaces, as formalized by the following equations:
\begin{equation}
    \mathbf{f}_a' = \text{FC}(\mathbf{f}_a) + \mathbf{e}_a,
    \label{eq:attr_embedding}
\end{equation}
\begin{equation}
    \mathbf{f}_s' = [\mathbf{f}_s^1 + \mathbf{e}_s, \mathbf{f}_s^2 + \mathbf{e}_s, \dots, \mathbf{f}_s^n + \mathbf{e}_s].
\label{eq:superpixel_embedding}
\end{equation}

Here, a learnable fully connected (FC) layer is used to project the attribute embeddings, explicitly ensuring their dimensionality is properly aligned with that of the superpixel token embeddings.
Leveraging the multi-head self-attention mechanism in Transformers \cite{vaswani2017attention}, we adopt an adapted architecture in which the attribute embedding $\mathbf{f}_a'$ explicitly serves as the query, while the superpixel token embedding $\mathbf{f}_s'$ simultaneously provides the corresponding key and value for each attention head. Scaled dot-product attention is then used to compute query-key affinities, enabling dynamic aggregation of attribute-aware superpixel features. Formally, the attentive representation $\mathbf{f}_{as}^i \in \mathbb{R}^d$ at the $i$-th head is defined as:
\begin{equation}
    \mathbf{f}_{as}^i = \text{softmax}\left(\frac{\mathbf{QK}^T}{\sqrt{d}}\right)\mathbf{V},
\label{eq:attention_head}
\end{equation}
where $\mathbf{Q} = \mathbf{f}_a' \mathbf{W}_i^q$, $\mathbf{K} = \mathbf{f}_s' \mathbf{W}_i^k$, and $\mathbf{V} = \mathbf{f}_s' \mathbf{W}_i^v$, with $\mathbf{W}_i^q, \mathbf{W}_i^k, \mathbf{W}_i^v \in \mathbb{R}^{D \times d}$ denoting the respective projection matrices for the query, key, and value.  
After computing all $h$ attention heads, their outputs are concatenated and passed through a final output projection layer to obtain the resulting attribute-aware representation:
\begin{equation}
    \mathbf{f}_{as} = [\mathbf{f}_{as}^1 \oplus \mathbf{f}_{as}^2 \oplus \dots \oplus \mathbf{f}_{as}^h]\mathbf{W}_3,
\label{eq:attr_aware_output}
\end{equation}
where $\mathbf{W}_3 \in \mathbb{R}^{hd \times D}$ denotes the output projection matrix.  
Following the standard Transformer architecture \cite{vaswani2017attention}, we augment $\mathbf{f}_{as}$ with an explicit residual connection by adding the mean-pooled superpixel token embeddings $\mathbf{f}_s'$. A subsequent multilayer perceptron (MLP) with residual connection and layer normalization is then applied to reliably generate the final attribute-aware image representation $\mathbf{f}_{\text{T}} \in \mathbb{R}^c$, which constitutes the output of the framework.
\subsection{Training and Inference}
The primary objective of training is to learn robust and discriminative attribute-aware representations, such that images sharing the same attribute values are represented with high similarity, while those with differing attribute values are explicitly and distinctly separated. To achieve this, we adopt a contrastive learning strategy incorporating both triplet loss and cross-InfoNCE loss. Specifically, we construct a triplet set $\mathcal{B} = \{(I_{i}, I_{i}^{+}, I_{i}^{-})\}_{i=1}^{N}$, where $I_{i}$ and $I_{i}^{+}$ share the same value with respect to the attribute $a$, whereas $I_{i}^{-}$ exhibits a different value. Here, $N$ denotes the mini-batch size. The attribute-related representation $\mathbf{f}^{i}_{\text{A}}$ and the attribute-aware representation $\mathbf{f}^{i}_{\text{T}}$ are learned separately via triplet ranking loss \cite{Veit_CSN_2017, Ma_ASEN_2020}. Formally, for a given mini-batch, the triplet loss is defined as:
\begin{equation}
\begin{aligned}
    &\mathcal{L}_{\text{A}} =\frac{1}{N}\sum_{i=1}^{N}\max(0,m-s(\mathbf{f}^{i}_{\text{A}},\mathbf{f}^{i+}_{\text{A}})+s (\mathbf{f}^{i}_{\text{A}},\mathbf{f}^{i-}_{\text{A}})), \\
    &\mathcal{L}_{\text{T}} =\frac{1}{N}\sum_{i=1}^{N}\max(0,m-s(\mathbf{f}^{i}_{\text{T}},\mathbf{f}^{i+}_{\text{T}}) +s (\mathbf{f}^{i}_{\text{T}},\mathbf{f}^{i-}_{\text{T}})).
\end{aligned}
\label{eq:triplet}
\end{equation}
Here, $\mathcal{L}_{\text{A}}$ denotes the loss for attribute-guided attention, while $\mathcal{L}_{\text{T}}$ corresponds to the loss for the superpixel token-based Transformer.  
$\mathbf{f}^{i}_{\text{T}}$, $\mathbf{f}^{i+}_{\text{T}}$, and $\mathbf{f}^{i-}_{\text{T}}$ represent the final attribute-aware representations of the $i$-th triplet $I_{i}$, $I_{i}^{+}$, and $I_{i}^{-}$ generated by \textit{Super}Fashion. Similarly, $\mathbf{f}^{i}_{\text{A}}$, $\mathbf{f}^{i+}_{\text{A}}$, and $\mathbf{f}^{i-}_{\text{A}}$ denote the corresponding attribute-related feature representations of the same triplet obtained from the attribute-guided attention module. The similarity function $s(\cdot, \cdot)$ is implemented as cosine similarity, and the margin hyperparameter $m$ enforces a minimum separation between positive and negative pairs. 
In addition, considering that attribute-related and attribute-aware features may share certain meaningful similarities yet also exhibit subtle differences, we aim to learn joint discriminative representations effectively from them. To this end, we propose a novel cross-InfoNCE loss $\mathcal{L}_{\text{AT}}$ to carefully align these two types of representations through cross-modal learning, formulated as follows:
\begin{table*}[!h]
\centering

  \caption{Comparative results (\%) on FashionAI dataset across each attribute and overall MAP metrics.} 
\label{tab:fashionai}
\renewcommand{\arraystretch}{1.0}
% \resizebox{\linewidth}{!}{
\begin{tabular}{ccccccccccc}
\toprule
\multirow{3}{*}{\textbf{Domain}} & \multirow{3}{*}{\textbf{Method}} & \multicolumn{8}{c}{\textbf{MAP for each attribute}} & \multirow[c]{3}{*}{\textbf{Overall MAP}}\\
\cmidrule(lr){3-10} 
 && skirt & sleeve & coat & pant & collar & lapel & neckline & neck &  \\
&& length & length & length & length & design & design & design & design & \\ 
\midrule
\multirow{7}{*}{\textbf{Prior SOTA}} & CSN \cite{Veit_CSN_2017}              & 61.97 & 45.06 & 47.30 & 62.85 & 69.83 & 54.14 & 46.56 & 54.47 & 53.52 \\
&ASEN  \cite{Ma_ASEN_2020}           & 64.44 & 54.63 & 51.27 & 63.53 & 70.79 & 65.36 & 59.50 & 58.67 & 61.02 \\
&HAEN \cite{yan2021learning}          & 64.13 & 55.52 & 56.41 & 72.31 & 73.32 & 69.22 & 62.41 & 59.80 & 64.13 \\
&AttnFashion \cite{wan2024learning}      & 65.70 & 56.46 & 54.64 & 71.12 & 74.45 & 69.36 & 65.69 & 65.54 & 65.37 \\
&ISLN    \cite{ISLN_2022}         & 65.91 & 58.83 & 56.45 & 71.22 & 74.53 & 70.55 & 65.71 & 65.61 & 66.10 \\
&ASEN++   \cite{Dong_ASEN++_2021}        & 66.34 & 57.53 & 55.51 & 68.77 & 72.94 & 66.95 & 66.81 & 67.01 & 64.31 \\
&RPF  \cite{Dong_RPF_2023}    & 66.75 & 67.84 & 59.59 & 73.14 & 75.72 & 73.18 & \textbf{74.40} & 74.98 & 70.10 \\ 
\midrule
\multirow{5}{*}{\textbf{SOTA-KD}} &ASEN\_V2+PKD \cite{xiao2024boosting} &69.28&62.13&59.72&73.08&\textbf{80.11}&74.08&68.98&70.04&68.48 \\
&ASEN\_V2+PT+PKD \cite{xiao2024boosting}&68.94&62.13&60.88&73.56&78.20&\underline{77.77}&69.94&69.32&69.14 \\
&ASEN+GeoDCL \cite{xiao2025geodcl}&65.20&53.95&50.42&67.10&76.32&70.47&64.60&67.55&62.81 \\
&ASEN\_V2+GeoDCL\cite{xiao2025geodcl} &68.71&59.18&55.54&70.72&77.14&73.03&68.49&69.25&66.48\\
&RPF+GeoDCL \cite{xiao2025geodcl} & \underline{69.96}&\underline{68.70}&\underline{61.05}&\underline{73.96}&78.34&77.19&\underline{70.72}&\underline{80.01}&\underline{71.15} \\
\midrule 
\textbf{Ours} & \textbf{\textit{Super}Fashion} &  \textbf{70.48} & \textbf{69.57} & \textbf{61.90} & \textbf{74.06} & \underline{79.82} & \textbf{78.12} & 70.39 & \textbf{80.19} & \textbf{72.46} \\ 
\bottomrule
\end{tabular}
 % }
\end{table*}
\begin{table*}[!h]
\centering

 \caption{Comparative results (\%) on DARN dataset across each attribute and overall MAP metrics.} 
\label{tab:darn}
\renewcommand{\arraystretch}{1.0}
\setlength{\tabcolsep}{4pt}
\begin{tabular}{cccccccccccc}
\toprule
\multirow{3}{*}{\textbf{Domain}} & \multirow{3}{*}{\textbf{Method}} & \multicolumn{9}{c}{\textbf{MAP for each attribute}} & \multirow[c]{3}{*}{\textbf{Overall MAP}} \\
\cmidrule(lr){3-11} 
&& clothes & clothes & clothes & clothes & clothes & clothes & collar & sleeve & sleeve &\\
&& category & button & color & length & pattern & shape & shape & length & shape & \\ \midrule
\multirow{7}{*}{\textbf{Prior SOTA}}& CSN \cite{Veit_CSN_2017} & 34.10 & 44.32 & 47.38 & 53.68 & 54.09 & 56.32 & 31.82 & 78.05 & 58.76 & 50.86 \\ 
&ASEN \cite{Ma_ASEN_2020} & 36.69 & 46.96 & 51.35 & 56.47 & 54.49 & 60.02 & 34.18 & 80.11 & 60.04 & 53.31 \\ 
&HAEN \cite{yan2021learning} & 32.10 & 47.04 & 45.03 & 48.27 & 49.92 & 51.22 & 28.05 & 78.29 & 58.47 & 48.70 \\ 
&AttnFashion \cite{wan2024learning} & 34.94 & 48.56 & 48.14 & 54.47 & 52.65 & 56.36 & 32.32 & 82.63 & 60.77 & 52.32 \\ 
&ISLN \cite{ISLN_2022} & 38.84 & 51.26 & 52.67 & 56.55 & 53.85 & 58.34 & 36.64 & 82.74 & \underline{61.28} & 54.68 \\ 
&ASEN++ \cite{Dong_ASEN++_2021} & 40.15 & 50.42 & 53.78 & 60.38 & \underline{57.39} & 59.88 & 37.65 & 83.91 & 60.70 & 55.94 \\ 
&RPF \cite{Dong_RPF_2023} & \underline{44.60} & \underline{55.30} & \underline{54.02} & \underline{63.85} & 56.91 & \underline{60.15} & \underline{38.70} & \underline{84.57} & 59.35 & \underline{56.88} \\ 
\midrule
\textbf{Ours}&\textbf{\textit{Super}Fashion} & \textbf{48.66} & \textbf{58.10} & \textbf{57.52} & \textbf{69.80} & \textbf{57.86} & \textbf{64.51} & \textbf{39.10} & \textbf{86.77} & \textbf{62.53} & \textbf{62.15} \\
\bottomrule
\end{tabular}
 % }
\end{table*}
\begin{equation}
    \mathcal{L}_{\text{AT}} = -\frac{1}{N} \sum_{i=1}^{N} \log \left( \frac{\mathcal{Z}^+}{\mathcal{Z}^+ + \mathcal{Z}^-} \right),
    \label{eq:loss}
\end{equation}
where $\mathcal{P}$ and $\mathcal{N}$ denote the positive and negative sets of the representations that share or differ in attribute values with $I_{i}$, respectively.
The partition functions $\mathcal{Z}^{+}$ and $\mathcal{Z}^{-}$ are defined as follows:  
\begin{equation}
\begin{aligned}
    &\mathcal{Z}^{+} = \exp(\mathbf{f}_{\text{T}}^i \cdot \mathbf{f}_{\text{A}}^i/ \tau)  +   \sum_{\mathbf{f}_{\text{A}}^{j+} \in \mathcal{P}} \exp(\mathbf{f}_{\text{T}}^i \cdot \mathbf{f}_{\text{A}}^{j+} / \tau), \\
    &\mathcal{Z}^{-} = \sum_{\mathbf{f}_{\text{A}}^{j-} \in \mathcal{N}} \exp(\mathbf{f}_{\text{T}}^i \cdot \mathbf{f}_{\text{A}}^{j-} / \tau).
\end{aligned}
\label{eq:lossAT}
\end{equation}

Consequently, the final overall loss function is expressed as:
\begin{equation}
    \mathcal{L} = \mathcal{L}_{\text{A}} + \alpha \mathcal{L}_{\text{T}} + \beta \mathcal{L}_{\text{AT}},
    \label{eq:finalloss}
\end{equation}
where $\alpha$ and $\beta$ are training hyperparameters that balance the contributions of the respective loss components. 

During inference, the similarity between a query image $I$ and a candidate image $I^*$ with respect to a specific attribute is computed as:
\begin{equation}
    \text{sim}(I,I^*) = \lambda s(\mathbf{f}_{\text{A}}, \mathbf{f}^{*}_{\text{A}}) + (1-\lambda) s(\mathbf{f}_{\text{T}}, \mathbf{f}^{*}_{\text{T}}),
    \label{eq:inference}
\end{equation}
where $\lambda$ is a weighting hyperparameter used during inference, and $s(\cdot,\cdot)$ denotes the similarity function, such as cosine similarity.

\section{Experiment}
\subsection{Experimental Setup}

\subsubsection{Datasets} 
To ensure a fair and rigorous comparison, following prior studies~\cite{Ma_ASEN_2020,Dong_ASEN++_2021,Dong_RPF_2023,Jiao_M3NET_2023,xiao2025geodcl}, we evaluate our proposed framework \textit{Super}Fashion on three widely used benchmark datasets: \textit{FashionAI} \cite{dataset_FashionAI}, \textit{DeepFashion} \cite{dataset_DeepFashion}, and \textit{DARN} \cite{dataset_DARN}. The dataset partitioning and preprocessing procedures are kept consistent with those adopted in previous works. 
It is worth noting that images in the \textit{DeepFashion} dataset are annotated with multiple attributes, whereas both the \textit{DARN} and \textit{FashionAI} datasets provide a single-attribute label for each image.
\begin{table*}[t]
\centering

\caption{Comparative results (\%) on DeepFashion dataset across each attribute and overall MAP metrics.} 
\label{tab:deepfashion}
\renewcommand{\arraystretch}{1.0}
% \resizebox{0.8\linewidth}{!}{
\setlength{\tabcolsep}{12pt}
\begin{tabular}{cccccccc}
\toprule
\multirow{2}{*}{\textbf{Domain}} &\multirow{2}{*}{\textbf{Method}} & \multicolumn{5}{c}{\textbf{MAP for each attribute}} & \multirow[c]{2}{*}{\textbf{Overall MAP}} \\
\cmidrule(lr){3-7} 
&&texture & fabric & shape & part & style &  \\
\midrule
% Triplet Network & 13.26 & 6.28 & 9.49 & 4.43 & 3.33 & 7.36 \\ 
\multirow{4}{*}{\textbf{Prior SOTA}} &CSN \cite{Veit_CSN_2017} & 14.09 & 6.39 & 11.07 & 5.13 & 3.49 & 8.01 \\ 
&ASEN \cite{Ma_ASEN_2020} & 15.01 & 7.32 & 13.32 & 6.27 & 3.85 & 9.14 \\ 
&AttnFashion \cite{wan2024learning} & 12.90 & 6.34 & 11.38 & 5.24 & 4.20 & 8.01 \\ 
&ASEN++ \cite{Dong_ASEN++_2021} & 15.60 & 7.67 & 14.31 & 6.60 & 4.07 & 9.64 \\ 
&RPF \cite{Dong_RPF_2023} & 15.62 & 8.30 & 15.02 & 7.38 & 4.77 & 10.22 \\ 
\midrule
\multirow{3}{*}{\textbf{SOTA-KD}} &ASEN+GeoDCL \cite{xiao2025geodcl}&16.09&7.84&12.80&6.27&5.25&9.41 \\
&ASEN\_V2+GeoDCL \cite{xiao2025geodcl}&15.29&7.11&11.77&5.52&3.76&8.68 \\
&RPF+GeoDCL \cite{xiao2025geodcl} &\underline{16.69}&\underline{8.95}&\underline{15.47}&\underline{8.02}&\underline{5.19}&\underline{10.80} \\
\midrule
\textbf{Ours}&\textbf{\textit{Super}Fashion} & \textbf{17.62} & \textbf{9.90} & \textbf{16.37} & \textbf{8.07} & \textbf{5.69} & \textbf{11.81} \\  
\bottomrule
\end{tabular}
% }
\end{table*}
\begin{table*}[htbp]
\centering

\caption{Cross-dataset evaluation results and performance for FashionAI $\rightarrow$ DARN and DARN $\rightarrow$ FashionAI settings. The notation S $\rightarrow$ T denotes training on dataset S and testing on dataset T. \textit{Italicized} results indicate in-dataset training and testing.}
\label{tab:cross-dataset}
\renewcommand{\arraystretch}{1.0}
% \resizebox{\linewidth}{!}{
\setlength{\tabcolsep}{5pt}
\begin{tabular}{ccccccccccc}
\toprule
\multirow[c]{3}{*}{\textbf{Method}} & \multicolumn{4}{c}{\textbf{FashionAI $\rightarrow$ DARN}} & \multicolumn{4}{c}{\textbf{DARN $\rightarrow$ FashionAI}} \\
\cmidrule(lr){2-5} \cmidrule(lr){6-9}
& sleeve length & clothes length & collar shape & \makecell{\textbf{Overall} \\ \textbf{MAP}} &  sleeve length & coat length & neckline design & \makecell{\textbf{Overall} \\ \textbf{MAP}} \\
\midrule
ASEN \cite{Ma_ASEN_2020} & 65.63 & 43.67 & 24.08 & 37.46 &  29.36 & 25.08 & 16.86 & 23.35 \\
ASEN++ \cite{Dong_ASEN++_2021} & 65.68 & 44.35 & 24.08 & 38.05 &  30.56 & 26.08 & 17.26 & 24.31 \\
RPF \cite{Dong_RPF_2023} & 66.14 & 44.87 & 23.62 & 38.81 &  34.93 & 27.96 & 20.89 & 26.09 \\
\midrule
\multirow[c]{2}{*}{\textbf{\textit{Super}Fashion}} & \textit{86.77} & \textit{69.80} & \textit{39.10} & \textit{64.80} &   \textit{69.57} & \textit{61.90} & \textit{70.39} & \textit{63.31} \\
& \textbf{67.55} & \textbf{46.76} & \textbf{27.30} & \textbf{41.57} &  \textbf{38.51} & \textbf{30.84} & \textbf{23.41} & \textbf{29.47} \\
\bottomrule
\end{tabular}
% }
\end{table*}
\subsubsection{Baseline Models} 
We compare our framework with a comprehensive set of representative SOTA methods that have been previously introduced and discussed in Sec.~\ref{sec:relatedwork}. These baselines encompass both earlier and more recent studies, including CSN~\cite{Veit_CSN_2017}, ASEN~\cite{Ma_ASEN_2020}, HAEN~\cite{yan2021learning}, ISLN~\cite{ISLN_2022}, ASEN++~\cite{Dong_ASEN++_2021}, AttnFashion~\cite{wan2024learning}, and RPF~\cite{Dong_RPF_2023}.
In addition to these foundations, we also incorporate more advanced frameworks such as GeoDC$\text{L}$~\cite{xiao2025geodcl}, 
which enhances knowledge distillation by enforcing geometric consistency on prior SOTA models, and PKD~\cite{xiao2024boosting}, 
which leverages progressive knowledge disentanglement to further improve existing approaches.
% \subsubsection{Implementation Details}
\subsubsection{Implementation Details} 
Consistent with prior works~\cite{Ma_ASEN_2020,Dong_ASEN++_2021,Dong_RPF_2023,Jiao_M3NET_2023,xiao2025geodcl}, we employ mean average precision (MAP) as the evaluation metric across all datasets, reporting MAP for each attribute as well as the overall MAP.
For the Transformer, we employ ViT-B/16 network pre-trained on ImageNet and employ ResNet50 pre-trained on ImageNet as the local feature encoder, owing to its effectiveness in capturing spatial structural information from images. Each attribute is represented as a one-hot ID and mapped to a learnable embedding that guides visual feature extraction, following common practice in ASFR.
The training procedure consists of two stages, consistent with~\cite{Dong_ASEN++_2021}:  
\begin{enumerate}
    \item The initial learning rate is set to $1 \times 10^{-4}$ and decays by a factor of 0.3 every three epochs, for a total of 50 epochs.  
    \item The learning rate is then reduced to $1 \times 10^{-5}$ and decays by a factor of 0.95 at each epoch, for an additional 50 epochs.  
\end{enumerate}
We set the hyperparameters as follows: $m = 0.2$ for Eq.~(\ref{eq:triplet}), $\tau = 0.07$ for Eq.~(\ref{eq:lossAT}), while $\alpha = 0.1$ and $\beta = 0.04$ are used in Eq.~(\ref{eq:finalloss}), and finally, $\lambda = 0.3$ is applied in Eq.~(\ref{eq:inference}) for all cases as well.
\subsection{Main Experimental Results and Analysis}  

\subsubsection{Comparison to Baseline Models}  
Overall, the comparison with baseline models demonstrates that \textit{Super}Fashion establishes new SOTA performance, delivering consistent and substantial improvements across multiple datasets for ASFR tasks. The gains are largely attributed to the integration of innovative superpixel segmentation with a superpixel token-based Transformer architecture.

As summarized in Tables \ref{tab:fashionai}–\ref{tab:deepfashion}, \textit{Super}Fashion consistently and significantly outperforms prior SOTA methods by a substantial margin in terms of overall MAP. Specifically, on the FashionAI and DeepFashion datasets, it achieves relative increases of \textbf{1.84\%} and \textbf{9.35\%}, respectively, compared with the previous leading method RPF+GeoDCL \cite{xiao2025geodcl}. Moreover, on the DARN dataset, it substantially surpasses the previous SOTA approach RPF \cite{Dong_RPF_2023} by a relative margin of \textbf{9.27\%}, further demonstrating the framework’s superior attribute-specific fashion retrieval capability and robust generalization.

In addition to its overall performance, \textit{Super}Fashion exhibits notable advantages in fine-grained attribute-specific fashion retrieval across heterogeneous feature distributions. For instance, on the DeepFashion dataset, the \texttt{texture} and \texttt{fabric} attributes show relative improvements of \textbf{5.57\%} and \textbf{10.61\%}, 
\begin{table*}[t]
    \centering
    
         \caption{Computational time efficiency comparison on FashionAI, DARN, and DeepFashion datasets.}
    \label{tab:time}
    \renewcommand{\arraystretch}{1.0}
    \setlength{\tabcolsep}{12pt}
    % \resizebox{\linewidth}{!}{
    \begin{tabular}{c c c c c}
    \toprule
    \textbf{Dataset} & \textbf{Method} & \textbf{Avg Time ($s$ / image) }& \textbf{Avg Time ($\mu s$ / image pair)} & \textbf{Throughput (QPS)} \\
    \midrule
    \multirow[c]{2}{*}{\textbf{FashionAI}}&RPF& 0.275 & 19.10 & 3.64 \\
    &\textbf{\textit{Super}Fashion} & 0.306 & 21.25 & 3.27  \\ \midrule
    \multirow[c]{2}{*}{\textbf{DARN}}&    RPF& 0.200 & 22.09 & 5.00 \\
    &\textbf{\textit{Super}Fashion} & 0.220 & 24.30 & 4.55  \\
    \midrule
     \multirow[c]{2}{*}{\textbf{DeepFashion}}&RPF& 0.105 & 4.75 & 9.52 \\
    &\textbf{\textit{Super}Fashion }& 0.121 & 5.48 & 8.26 \\ 
    \bottomrule
    \end{tabular}
    % }
\end{table*}
respectively, over the previous SOTA RPF+GeoDCL \cite{xiao2025geodcl}, 
highlighting the framework’s capability to capture intricate micro-structural patterns. These improvements are further emphasized by the observed ability of \textit{Super}Fashion to handle diverse and challenging attribute variations in fashion data. Similarly, on the DARN dataset, the \texttt{clothes category} and \texttt{clothes button} attributes experience gains of \textbf{9.10\%} and \textbf{5.06\%}, respectively, compared to RPF \cite{Dong_RPF_2023}. 
Additionally, on DARN dataset, \textit{Super}Fashion achieves relative improvements of \textbf{6.48\%} and \textbf{9.32\%} on the \texttt{clothes color} and \texttt{clothes length} attributes, respectively, underscoring its robustness and versatility in capturing discriminative features of local attributes.
This consistent trend of improvement across different datasets demonstrates the framework’s broad applicability to various fashion-related tasks.
\subsubsection{Cross-Dataset Generalization} 
\label{sec:cross}

\textit{Super}Fashion exhibits robust cross-dataset knowledge transfer and generalization capabilities. To evaluate this property, we assess its performance on corresponding attributes across the DARN 
and FashionAI datasets, despite differences in attribute values. In this setting, the attributes \texttt{sleeve length}, \texttt{coat length}, and \texttt{neckline design} in the FashionAI dataset correspond to \texttt{sleeve length}, \texttt{clothes length}, and \texttt{collar shape} in the DARN dataset, respectively. As shown in Table \ref{tab:cross-dataset}, \textit{Super}Fashion consistently and significantly outperforms the previous SOTA method RPF \cite{Dong_RPF_2023} in cross-dataset transfers, achieving relative improvements in overall MAP of \textbf{7.11\%} when transferring from FashionAI to DARN, and \textbf{12.96\%} when transferring from DARN to FashionAI. These results clearly underscore the framework’s superior knowledge transfer capability and its exceptional generalization performance across heterogeneous datasets.

\begin{table}[t]
\centering

 \caption{Ablation results for the different contributions of \textit{Super}Fashion's key components on DeepFashion dataset.}
\label{tab:ablation}
\renewcommand{\arraystretch}{1.0}
\setlength{\tabcolsep}{3pt}
\begin{tabular}{ccccccc}
\toprule
\multirow{2}{*}{\textbf{Method}} & \multicolumn{5}{c}{\textbf{MAP for each attribute}} & \multirow[c]{3}{*}{\makecell{\textbf{Overall }\\ \textbf{MAP}}}\\
\cmidrule(lr){2-6} &texture & fabric & shape & part & style& \\ \midrule
w/o Attention &15.22 & 8.57 & 15.28 & 7.61 & 4.23 & 10.28\\ 
w/o Transformer &15.01 & 8.23 & 14.96 & 7.01 & 3.89 & 9.65\\ 
\textbf{\textit{Super}Fashion} & \textbf{17.62} & \textbf{9.90} & \textbf{16.37} & \textbf{8.07} & \textbf{5.69} & \textbf{11.81} \\  
\bottomrule
\end{tabular}

\end{table}

\subsection{Time Efficiency Analysis}
\label{sec:time}
The throughput of \textit{Super}Fashion experiences a slight reduction compared to RPF \cite{Dong_RPF_2023}, but this decrease is minor when weighed against the substantial performance gains. We conduct a detailed evaluation of \textit{Super}Fashion's computational time efficiency, as summarized in Table \ref{tab:time}. Specifically, the framework exhibits a modest throughput decrease of 0.37 QPS, 0.45 QPS, and 0.26 QPS on the FashionAI, DARN, and DeepFashion datasets compared with RPF \cite{Dong_RPF_2023}, respectively. This reduction arises primarily from the additional computational overhead incurred by superpixel generation during inference. Importantly, these costs are more than compensated by notable improvements in overall MAP, with relative \textbf{gains} of 3.37\%, 9.27\%, and 15.56\% across the same datasets over RPF. These results indicate that \textit{Super}Fashion achieves a favorable balance between computational efficiency and retrieval performance, demonstrating its practicality and effectiveness for real-world ASFR tasks.
\subsection{Ablation Study}
\subsubsection{Effect of Key Component}
Table \ref{tab:ablation} illustrates the performance variations of our framework when key components, namely the attribute-guided attention mechanism or the superpixel token-based Transformer, are removed. Ablation of either component results in a substantial performance decline. Notably, the removal of the superpixel token-based Transformer leads to the most significant drop, as it impairs the framework’s ability to effectively capture fine-grained, micro-structural attribute patterns.
\begin{table}[t]
    \centering
    
        \caption{Ablation results on three datasets for the choice of key component across the overall MAP evaluation metric.}
    \label{tab:keychoice}
    \renewcommand{\arraystretch}{1.0}
    % \resizebox{0.9\linewidth}{!}{
    \setlength{\tabcolsep}{8pt}
    \begin{tabular}{cccc}
    \toprule
    \textbf{Method}& \textbf{FashionAI} & \textbf{DARN} & \textbf{DeepFashion} \\ \midrule
    Patch Token &70.05&56.93&10.24\\ 
    SLIC &71.89&61.47&11.01\\ 
    \textbf{\textit{Super}Fashion} &\textbf{72.46}&\textbf{62.15}&\textbf{11.81}\\
    \bottomrule
    \end{tabular}
    % }
\end{table}
\subsubsection{Choice of Key Component}
As presented in Table \ref{tab:keychoice}, replacing superpixel tokens with conventional $16 \times 16$ patch tokens, as utilized in RPF \cite{Dong_RPF_2023}, results in a substantial performance decline, with relative \textbf{reductions} in overall MAP of \textbf{3.33\%}, \textbf{8.40\%}, and \textbf{13.29\%} across FashionAI, DARN, and DeepFashion datasets, yielding an overall MAP only marginally above that of RPF \cite{Dong_RPF_2023}. In contrast, substituting the superpixel generation algorithm with SLIC \cite{achanta2012slic} induces only minor performance variations. While employing more sophisticated superpixel generation algorithms or models could potentially enhance performance further, resource constraints must be taken into account.
\begin{figure*}[htbp]
    \centering
    \includegraphics[width=\linewidth]{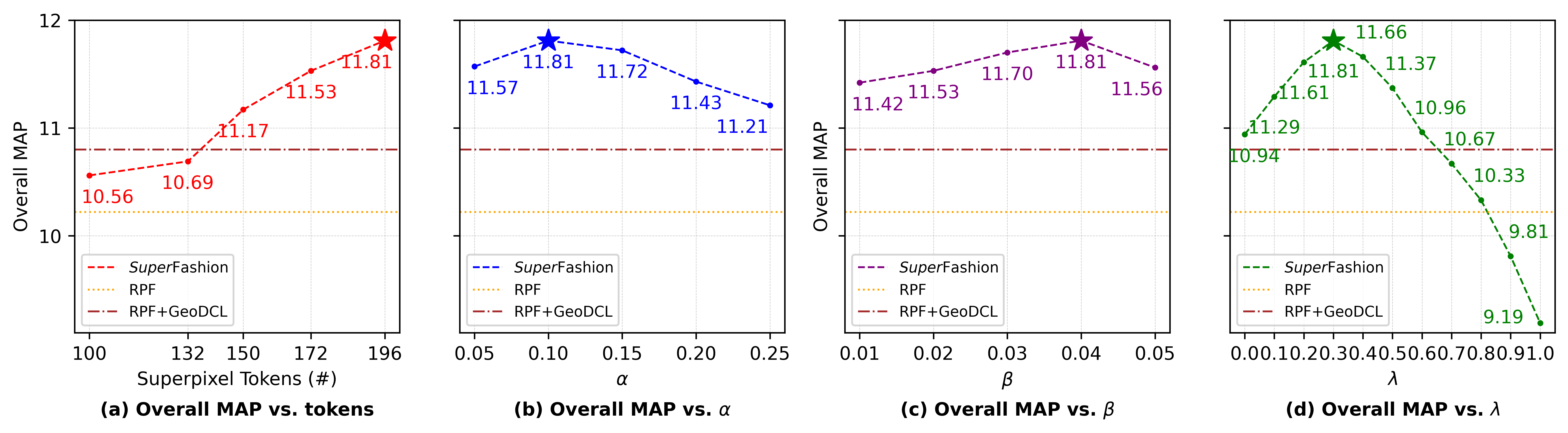}

    \caption{Overall MAP vs. superpixel tokens and hyperparameters $\alpha, \beta, \lambda$ on the DeepFashion dataset.}
    \label{fig:super}
\end{figure*}
% \vspace{-2pt}
\begin{figure}[!h]
    \centering
\includegraphics[width=\linewidth]{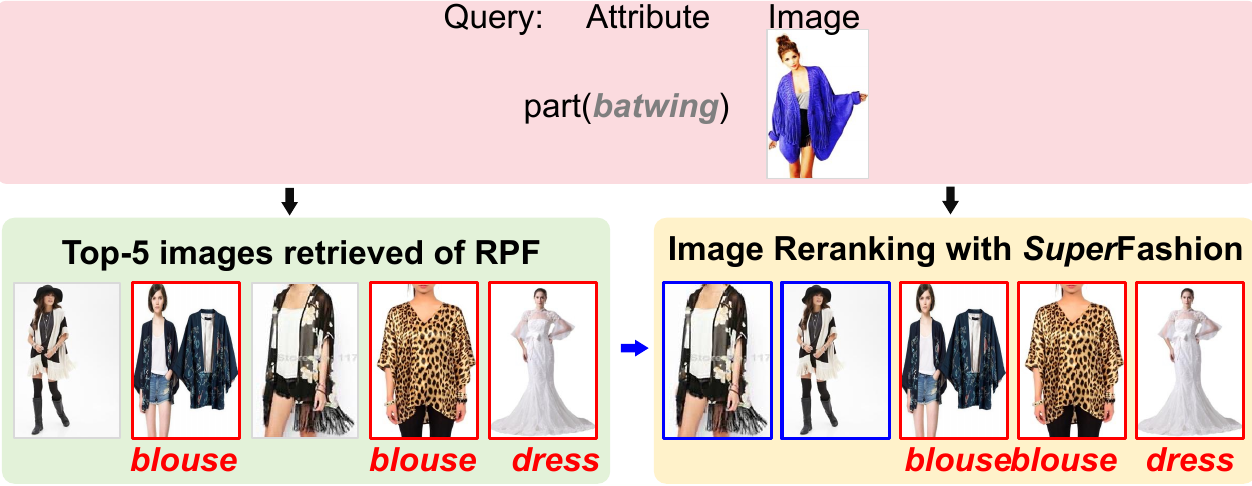}

    \caption{Retrieval case with incorrect retrievals highlighted.}
    \label{fig:case_study}
\end{figure}
\subsubsection{Impact of Superpixel Token Count}  
Figure \ref{fig:super} (a) depicts the influence of varying superpixel token counts on the performance of \textit{Super}Fashion. As the number of superpixel tokens increases, the framework exhibits a consistent and notable performance enhancement. This improvement is likely attributable to the framework's enhanced capability to capture intricate, fine-grained micro-structural patterns of attributes with higher token counts.
\subsection{Hyperparameter Analysis} 
We conduct an study on the DeepFashion dataset to assess the impact of hyperparameters $\alpha$, $\beta$, and $\lambda$, with results shown in Figure~\ref{fig:super} (b)-(d). The results reveal that $\alpha$ and $\beta$ show low sensitivity over the ranges 0.05-0.25 and 0.01-0.05, respectively, while $\lambda$ is highly sensitive, with MAP peaking at $\lambda=0.3$. The parameter $\lambda$ balances attribute-related and attribute-aware features during inference, with optimal performance observed for $\lambda$ in the range 0.0--0.5, outperforming $\lambda$ values of 0.6--1.0. This is attributed to the Superpixel Token-based Transformer, which enhances fine-grained attribute feature learning after noise-irrelevant features are filtered by attribute-guided attention. Notably, performance at $\lambda=0$ surpasses that at $\lambda=1$, reinforcing the effectiveness of the superpixel token-based Transformer, consistent with ablation study results.
\subsection{Case Study}
\subsubsection{Retrieval Case} 
We present example cases of the ASFR task, comparing \textit{Super}Fashion with the baseline RPF \cite{Dong_RPF_2023}. For each query image and specified attribute, the top five retrieved images are displayed. As shown in Figure~\ref{fig:case_study}, \textit{Super}Fashion effectively captures subtle, fine-grained attribute differences, whereas RPF often produces mismatches. 
\subsubsection{Visualization Analysis} 
\begin{figure}[!h]
    \centering
    \includegraphics[width = \linewidth]{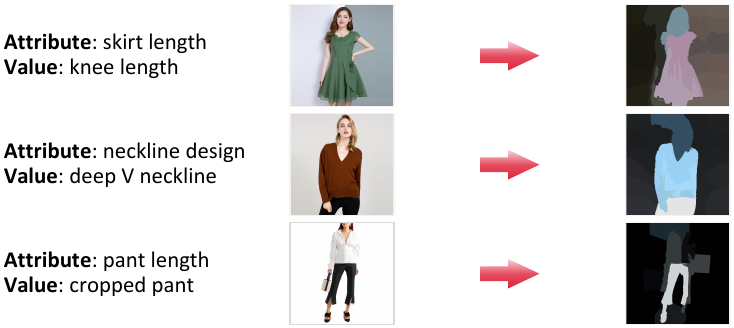}
    \caption{Attribute-based superpixel segmentation map.}
    \label{fig:vis}
\end{figure}

Figure \ref{fig:vis} presents visualization examples of attribute-based superpixel segmentation results. For the attributes \texttt{skirt length}, \texttt{neckline design}, and \texttt{pant length}, the segmentation highlights the relevant attribute-specific features and structures. These results provide robust support for \textit{Super}Fashion's ability to extract superpixel tokens for effective training.
\vspace{-15pt}
\section{Conclusion}% 
In this paper, we propose \textit{Super}Fashion, a novel framework for attribute-specific fashion retrieval. The framework first extracts attribute-related features to guide the cropping of meaningful image regions, and then generates compact, semantically coherent superpixel tokens, which are subsequently aggregated and processed by a superpixel token-based Transformer for adaptive feature interaction and feature fusion. Extensive experiments on multiple datasets clearly demonstrate that \textit{Super}Fashion effectively captures fine-grained attribute microstructures, mitigates background noise, and significantly outperforms existing SOTA methods. The results highlight the critical importance of semantically coherent tokenization for enhancing attribute-specific retrieval. 
For future work, we plan to explore attribute-guided superpixel segmentation, leveraging pre-trained attribute recognition models to provide semantic cues for precise alignment with attribute boundaries, thereby producing purer, more discriminative features and extending the framework to Web-based attribute-related retrieval across different domains.
\section{Acknowledgments}
This work was supported by the National Natural Science Foundation of China (Nos. 62406319, 62572465) and the Youth Innovation Promotion Association of CAS (No.2021153).
\balance
\bibliographystyle{ACM-Reference-Format}
\bibliography{ref}

\end{document}